\documentclass[a4paper,twoside]{article}
\usepackage{graphicx}
\usepackage{jcmse}
\usepackage{epsfig}

\newcommand{\QED}{\hspace*{\fill}\rule{2.5mm}{2.5mm}}

\def\AAA{\AA$^{3}$}

\newcommand{\JCMSEtitle}%
 {Polarizabilities of Intermediate Sized Lithium Clusters From Density-Functional Theory}

\newcommand{\JCMSEauthor}
  {\textbf{Rajendra R. Zope\footnote{rzope@utep.edu}}\\[1ex]
Department of Physics, University of Texas El Paso, El Paso TX, 79968,
and NSF CREST Center for Nanomaterials Characterization Science and Process
Technology, Howard University, School of Engineering, N.W. Washington D.C. 
20059 \\[1ex]\textbf{Tunna Baruah \footnote{tbaruah@utep.edu}}\\[1ex]
Department of Physics, University of Texas El Paso, El Paso TX 79968\\[1ex]
\textbf{Mark R. Pederson\footnote{mark.pederson@nrl.navy.mil}\footnote{Corresponding
author}}\\[1ex]Center for Computational Materials Science, Code 6392, Naval Research Laboratory, Washington DC 20375}

\newcommand{\yearnumber}{2}
\newcommand{\volumenumber}{2}
\newcommand{\issuenumber}{3}
\newcommand{\firstpage}{1}
\newcommand{\lastpage}{3}
\begin{document}

\maketitle \markboth{\JCMSEauthorshort{R.R. Zope et. al.}}%
{\JCMSEtitleshort{Polarizabilities of Li Clusters from NRLMOL/DFT}}

\centerline{Received 30 May 2007}

\begin{abstract}

   We present a detailed investigation of static dipole polarizability of lithium clusters
containing up to 22 atoms. We first build a database of lithium clusters by optimizing 
several candidate structures for the ground state geometry for each size. 
The full polarizability tensor is determined for about 5-6 isomers of each cluster size using 
the finite-field method. All calculations are performed using  large Gaussian basis sets,
and within the generalized gradient approximation to the density functional theory,
as implemented in the NRLMOL suite of codes. 
The average polarizability per atom varies from 11 to 9 \AAA\, within the 8-22 size range
and 
show smoother decrease with increase in cluster size 
than the experimental values.
While the average polarizability exhibits a relatively weak 
dependence on cluster conformation, significant changes in the degree of anisotropy of
the polarizability tensor are observed. 
Interestingly, in addition to the expected even
odd (0 and 1 $\mu_B$) magnetic states, our results show several cases where clusters with
an odd number of Li atoms exhibit elevated spin states (e.g. 3 $\mu_B$).

\end{abstract}



\begin{MSC}
\end{MSC}

\section{Introduction}
Materials composed purely of alkali atoms are expected to closely mimic the free-electron gas
or jellium models as they have only one very delocalized electron outside of a noble-gas 
shell.  Because of the relative  simplicity of pure alkali systems they are often viewed as
good systems for benchmarking quantum-mechanical methods and for investigating the transition
from localized to intinerent electronic behavior~\cite{mrp001,mrp003}. 
This is especially true for density-functional-based
calculations since the original approximations to density-functional theory~\cite{HK} were based upon
analytical expressions derived from exchange-only treatments of the the free-electron gas. For
example Slater derived an expression for a local potential which reproduced
the Hartree-Fock trace of the free electron gas~\cite{Slater}. 
Later, Kohn and Sham derived an expression that reproduced the Fermi
energy of the free electron gas~\cite{KS}. The latter expression 
also reproduces the total exchange-energy of a free-electron gas (See for example Ref.~\cite{mrp004}). 
The factor of
$3/2$ difference in the Kohn-Sham and Slater approximations also lead to the use of the X$_\alpha$ 
approximations~\cite{Slater,Xalpha} and approximations along the lines of this 
method continue to be investigated as an attractive means for developing
approximate element dependent functionals that permit fully analytic 
implementation of density functional theory~\cite{ZD_A,ZD_B}. 
Such analytic implementation is computationally very efficient and has 
been used  for structure optimization of icosahedral fullerenes 
containing more than two thousand atoms using triple-zeta quality 
basis set~\cite{ZD_C,ZD_D}. 

The ability to accurately reproduce polarizabilities of large systems is of significant importance
to many forefront research areas in computational chemistry, materials science and quantum physics.
The ability to accurately determine polarizabilities is required to account for hydrogen bonding, 
van der Waal's interactions,~\cite{mrp024,mrp025} solvation effects,~\cite{mrp026} and a materials dielectric response. As discussed
in Ref.~\cite{mrp024}, the same interactions or matrix elements required for an accurate determination
of phenomena that are directly mediated by polarizabilities also determine a variety of transition 
rates which include spontaneous emission, stimulated absorption and emission, and Foerster-energy 
transfer rates.  
Such rates are of direct importance to the problem of many photovoltaic 
applications. The radiative transition rates must also be quantified for  
applications to quantum-control of matter 
or any type of light-mediated manipulation of molecular- and cluster- materials. 
Derivatives of the polarizability tensor with respect to normal-mode displacement
also determine the intensity of the Raman shifts of a given molecule or 
cluster~\cite{mrp013,mrp017}.   

Because of the central role that polarizabilities play in materials science and chemistry there
have been significant efforts aimed at experimentally validating theoretically predicted 
polarizabilities. However, such experiments are themselves very difficult to interpret for
a variety of reasons. From the standpoint of comparison to experiments on
bulk systems, the polarizability of an array of nonoverlapping polarizable
molecules may be approximated from the Clausius-Mossotti relation. Such 
comparisons have been performed with some success on very idealized
systems such as fullerene molecules~\cite{mrp019,mrp010,mrp011,mrp012}. 
However, for pure-metal clusters the 
polarizability may not be determined from such a means because the individual clusters would
coalesce into the the bulk material if placed upon a lattice. As such the preferred experimental
approach for measurement of metal-cluster polarizabilities rely upon the electrostatic 
equivalent of a Stern-Gerlach experiment in which a beam of metal clusters
traverse a nonuniform electric field and are deflected due to the the
induced polarization of the cluster~\cite{Bonin,deHeer,Kresin,Knick,Li_expt}.
Again, for simple systems such a fullerene molecules quantitative agreement 
between theory and experiment
has been achieved. However, for metallic clusters deviations exist between different experiments
and also between experiment and theory. Generally, it appears that the theoretically predicted
polarizabilties display a more monotonic and smooth behavior than the experimental measurements.
Moreover, earlier comparisons suggested that the experimental polarizabilities tended
to be larger than theory. Such discrepancies are now largely understood to be due to temperature
dependent corrections that depend upon the permanent dipole of the clusters. However, some
discrepancies still exist and it is not understood why. One possibility that we attempt to 
investigate here is that different low-lying geometries of a cluster may have significantly 
different polarizabilites or anisotropies in their polarizabilities.

  Experimental measurements of polarizability of Li clusters containing up to 22 atoms 
have been reported~\cite{Li_expt}. Here, we present and compare our density functional 
predictions of  polarizabilities of Li clusters with the avialable experimental 
data. The calculations are 
performed for clusters in the size range 8-22 as number of 
studies have already addressed polarizabilities up to size 8~\cite{Maroulis_A,Pecul,Ghosh}.
Rather than concentrate on
the {\em ground state} for each size regime we have adopted a different approach that relies
on the generation of several low-energy structures. The polarizability of
each structure is subsequently calculated to determine whether conformer-induced changes in polarization are expected
to be observable.

\section{Computational Method}
 The determination of lowest energy structure of clusters is a  nontrivial task.
To obtain the candidate structures for the ground state of Li clusters, we 
make use of a recently determined database of sodium clusters~\cite{BZ_A}.
This database is generated using a  global optimization technique 
such as the basin hopping algorithm using the interatomic potential 
determined within the Kohn-Sham DFT framework. This numerical KS DFT 
provides very good description of sodium clusters~\cite{BZ_B,BZ_C,AMV}.
%
Using starting geometries from  this database and
Bachelet-Hamman-Schluter pseudo potential method ~\cite{BHS} we 
quickly generate low-energy structures for Li clusters using 
conjugate-gradient technique. A set of few lowest-energy geometries 
are then selected for further optimization at the 
all-electron level until forces on all atoms are below the 0.001 Hartree/Bohr. 
For each of these equilibrium 
geometries we have then performed six additional calculations with electric
fields turned on. The included fields are (E,0,0), (-E,0,0), (0,E,0), 
(0,-E,0), (0,0,E) and (0,0,-E) with E=0.001 atomic units. From the 
standard relation between the dipole moment, polarizability,
and applied electric field:
\begin{equation}
{p_i} = {\bf p_i^o} + \Sigma_j \alpha_{ij} E_j,
\end{equation}
we determine the polarizability tensor using a central differencing method. Additional discussion of 
this procedure may be found in Ref.~\cite{mrp013}. From the above expression it is possible to determine the 
central-difference off-diagonal elements ($\alpha_{ij}$) from either $p_i$ and $E_j$ or $p_j$ and $E_i$.
In the limit of a vanishingly small-field or an exactly harmonic polarizability tensor, the values
obtained in the two different ways would agree perfectly (regardless of issues related to basis sets).
This exact relation provides one way of estimating whether the applied electric field is small
enough to accurately determine the polarizability tensor. We find that the 
off-diagonal elements are small but agree to approximately 1-5\%. Of course,
the size of the off-diagonal element depends on the original choice of 
coordinate system. Our original coordinate system uses the principal 
axes of inertia. 
To determine the principal polarizability axes we then diagonalize
the polarizability tensor and rotate the molecule so that the molecular 
x, y and z axes coincide with the predicted principle polarizability axes. 
As a final check we have ascertained for several cases that the 
predicted polarizabilities are accurate and uncontaminated by higher-order
polarizabilities. For these tests, we perform an analysis similar to 
that of Quong and Pederson for the C$_{60}$ and Benzene 
polarizabilites~\cite{mrp010,mrp011,mrp012}. That is, we have determined that the polarizabilities obtained from either
energies or dipole moments agree.  

To perform the electronic-structure calculations we have used the NRLMOL suite of codes due to
Pederson {\em et al.}~\cite{mrp005,mrp006,mrp007,mrp015,mrp016}.  This all-electron method uses Gaussian-type orbitals to represent both 
the core, valence and unoccupied electrons. A highly accurate variational integration mesh~\cite{mrp005} is used
to calculate matrix elements required for the secular equation. The coulomb potential is calculated analytically
(in terms of incomplete gamma functions)~\cite{mrp016} on this mesh. The exchange-correlation potential and
energy kernel is also calculated analytically on this mesh. For the work discussed here we use the 
Perdew-Burke-Ernzerhof~\cite{PBE}
generalized gradient approximation. The basis sets here have been specifically optimized for this functional.
For Li, we use a total of ten single-gaussians to construct 5 contracted s-type orbitals, 3 contracted p-type
orbitals and 1-contracted d-type orbital. Each d-type orbital used also contributes a spherical 
s-type orbital
with an $r^2$ prefactor. This basis set is used for optimizing the geometry. 
\begin{table}
\begin{center}
\caption{Gaussian decay parameter ,$\alpha$ in bohr$^{-2}$, and 
contraction coefficients,c(1s) and c(2s), used
in this work. As discussed in the text, the default basis set also contains 
single gaussians. For the s functions we use 
$\alpha_8$-$\alpha_{10}$. 
The p functions use $\alpha_7$-$\alpha_{9}$.  The d functions use 
$\alpha_7$. For the polarizability calculations, we also include $\alpha_{10}$
and $\alpha_{8}$ for the p- and d- functions respectively.  
}
\label{tab:bs}
\begin{tabular}{|r|r|r|} \hline 
      $\alpha$  &    c(1s)     &      c(2s)   \\
     3200.240723&      0.087121&     -0.015568\\
      477.961517&      0.163030&     -0.029089\\
      108.464722&      0.278823&     -0.050146\\
       30.622913&      0.432000&     -0.078375\\
        9.925916&      0.565224&     -0.107102\\
        3.488695&      0.560806&     -0.116751\\
        1.282134&      0.363596&     -0.098758\\
        0.466942&      0.092471&     -0.040081\\
        0.076048&      0.001033&      0.053626\\
        0.028278&     -0.000155&      0.028178\\
\hline
\end{tabular}
\end{center}
\end{table}
The gaussian exponents ($\beta_i$) used here range between $3.3x10^4$ and 0.028 bohr$^{-2}$. As discussed in detail by Porezag
and Pederson, these exponents are optimized iteratively by performing an SCF calculation on the 
isolated atoms.~\cite{mrp015} An application of the Hellmann-Feynman theorem allows us to determine the derivative of the total energy with
respect to each decay parameter ($\beta_i$) and then optimize these parameters using the conjugate gradient
algorithm. For calculations on Raman spectra and polarizabilities, it is often necessary to include additional
polarization functions to account for the response of the electrons to the applied fields. In this work we
have included one additional p-type and 1 additional d-type wavefunction. With all of these considerations, the
final basis set used for these calculations has 5-s type, 2 $r^2$ s-type, 3x4 p-type,
and 5x2 d-type functions
or 29 basis functions per atom. The basis sets are displayed in Table 1. Calaminici {\em et al.}~\cite{cal1,cal2,cal3} have suggested that another appealing way of accounting for polarization
is to develop functions from the original set that are 
multiplied by x,y, and z respectively. We have also calculated polarizabilities using this scheme. 
These are compared with NRLMOL basis set used in present 
calculations in Table \ref{tab:ZC}. Agreement between 
two datasets is excellent.

\begin{table}
\begin{center}
\caption{Average polarizability (\AA$^3$) of Li clusters using 
the NRLMOL default basis-sets and a basis-set
generated using the augmentation method based on the method of  Calaminici {\em et al}.
Agreement between the two sets is excellent.}
\label{tab:ZC}
\begin{tabular}{|c|c|c|} \hline 
     Cluster &   NRLMOL   &   ZC       \\
\hline
        Li8-a  & 84.77   &    84.77   \\
        Li8-b  & 83.31   &    83.3     \\
        Li9-a  & 103.15  &    103.14  \\
        Li9-b  & 96.55   &    96.46   \\
        Li9-c  & 96.15   &    96.06   \\
        Li9-d  & 94.26   &    94.14   \\
        Li9-e  & 93.97   &    93.85   \\
        Li10-a & 103.9   &    103.85   \\
        Li10-b & 102.84  &    102.78  \\
        Li10-c & 104.89  &    104.84  \\
        Li10-d & 104.81  &    104.74  \\
        Li10-e & 106.93  &    106.87  \\
\hline
\end{tabular}
\end{center}
\end{table}
As a final check about the quality of basis set used for description 
of Li clusters polarizabilities, we calculated the polarizability 
of Li10-a cluster using very large basis of single optimized gaussians 
(10 s-type, 10 r$^2$ s-type, 10x3 p-type, and 10x5 d-type or 100 basis
functions per atom). This yields an average polarizability of 103.8 \AAA\, 
while default NRLMOL basis gives 103.9 \AAA. Thus, any       
uncertainties due to basis sets are quite negligible.

The method used for generating starting geometries and low-energy structures typically leads to many
replicas of the lowest energy geometries. Filtering these replicas has been performed in several stages. Each geometry
in this work is labeled by the number of atoms (N) and a letter indicating the starting
point of the structure. Because of 
the fact that there are multiple replicas in some cases the letters are not necessarily sequential. Identifying 
replicated clusters is a difficult task. In this work we have used the principle axes of the polarizability 
tensor to standardize the coordinate system for each conformer. In addition to being able to compare the
values of the polarizabilities, dipole moments and HOMO-LUMO energies, we can view these geometries sequentially
which further aids in identifying replicated structures.  To more carefully automate this procedure it seems that
a vibrational analysis, which includes a secondary inverse-hessian-based quenching to the equilibrium geometry should
be performed for each cluster. In addition to being able to compare the infrared and Raman spectra of the clusters, this
would lead to a final set of structures with atomic forces that are 10-100 times smaller than those obtained here. 
\begin{table}[!h]
\begin{center}
\caption{Calculated results as a function of cluster size and conformation for N=8 to N=11. Results
include moment, $\mu$ in Bohr magnetons, binding energy (B.E.) in eV, average and principle 
polarizabilities ($\langle \alpha \rangle$, 
$\alpha_{xx}$,
$\alpha_{yy}$,
$\alpha_{zz}$ in \AA$^3$, the dipole moment ($p_x, p_y, p_z$) in atomic units
and the HOMO ($\epsilon_H$),  LUMO ($\epsilon_L$)
energies in eV and HOMO-LUMO gap ($\Delta$) in eV respectively. 
The 
anisotropy of the polarizability tensor may be determined from the principle
polarizabilities (See Eq. 7 Ref.~\cite{mrp013}).
}
\begin{tabular}{|c|c|c|c|ccc|ccc|ccc|} \hline 
N & $\mu$ & B.E. (eV)& $\langle \alpha \rangle$&  $\alpha_{xx}$& $\alpha_{yy}$&$\alpha_{zz}$ & $p_x$& $p_y$& $p_z$&$\epsilon_H$&$\epsilon_L$&$\Delta$\\
\hline \hline

 8-a&  0&     -0.94&  84.77&  91.88&  81.31&  81.12&   0.01&   0.00&   0.00&  -3.09&  -2.24&   0.84\\
 8-b&  0&     -0.94&  83.31&  89.82&  80.08&  80.02&   0.00&   0.00&   0.00&  -3.10&  -2.23&   0.87\\
\hline
 9-a&  1&     -0.95& 103.15& 130.60&  92.06&  86.80&  -0.02&  -0.01&  -0.01&  -2.62&  -2.12&   0.50\\
 9-b&  1&     -0.95&  96.55& 117.00&  94.60&  78.04&   0.00&   0.00&   0.00&  -2.57&  -2.27&   0.30\\
 9-c&  1&     -0.96&  96.15& 109.26& 100.89&  78.29&   0.00&   0.00&   0.00&  -2.54&  -2.25&   0.28\\
 9-d&  1&     -0.96&  94.26& 118.37&  82.25&  82.15&   0.07&   0.04&  -0.04&  -2.32&  -1.93&   0.39\\
 9-e&  1&     -0.96&  93.97& 118.36&  81.85&  81.70&   0.06&  -0.03&   0.06&  -2.33&  -1.93&   0.39\\
\hline
10-a&  0&     -0.98& 103.90& 140.20&  89.92&  81.58&  -0.01&   0.00&  -0.01&  -2.87&  -2.26&   0.61\\
10-b&  0&     -0.98& 102.84& 138.37&  89.15&  81.01&   0.00&   0.00&   0.00&  -2.88&  -2.27&   0.61\\
10-c&  0&     -0.98& 104.89& 142.08&  90.50&  82.07&   0.00&   0.00&   0.00&  -2.86&  -2.20&   0.66\\
10-d&  0&     -0.98& 104.81& 142.19&  90.08&  82.14&  -0.01&   0.00&   0.01&  -2.85&  -2.19&   0.66\\
10-e&  0&     -0.99& 106.93& 146.14&  87.42&  87.22&   0.00&   0.00&   0.00&  -2.84&  -2.07&   0.78\\
\hline
11-a&  1&     -0.97& 114.89& 141.63& 116.78&  86.25&  -0.01&   0.00&   0.01&  -2.91&  -2.53&   0.38\\
11-b&  1&     -0.99& 117.89& 160.30& 106.19&  87.18&   0.00&   0.00&   0.00&  -2.78&  -2.24&   0.54\\
11-c&  1&     -0.99& 116.62& 158.24& 106.56&  85.05&   0.11&  -0.05&   0.11&  -2.83&  -2.35&   0.48\\
11-d&  1&     -1.00& 115.46& 157.40& 105.70&  83.26&   0.00&   0.00&   0.00&  -2.87&  -2.38&   0.50\\
11-e&  1&     -1.00& 115.46& 157.40& 105.70&  83.26&   0.00&   0.00&   0.00&  -2.88&  -2.39&   0.49\\
11-f&  1&     -1.00& 115.43& 157.33& 105.74&  83.23&  -0.01&   0.00&   0.00&  -2.88&  -2.39&   0.49\\
\hline \hline
\end{tabular}
\end{center}
\end{table}

In this work we have allowed for the possibility of spin-polarized solutions. For the clusters containing
an even number of electrons we have started the calculations with a net total moment and found that 
they converge to completely unpolarized solutions. The same thing has been done for the clusters containing
an odd number of electrons. In all cases except for one, the odd-atom clusters have converged to structures
with a net moment of 1 $\mu_B$ (e.g S=1/2). However, for the 17-atom clusters, we have found that some of the 
geometries prefer a net moment of 3 $\mu_B$.  In this work we have not investigated the possibility of 
antiferromagnetic spin ordering. Based on what is known about both Li$_2$ dimers and solid lithium, it is
unlikely that an antiferromagnetic arrangement of localized spins would be preferred for any of these
structures. For example, in Ref.~\cite{mrp001} the large-bond length antiferromagnetic state of the Li$_2$
molecule has been investigated within both density-functional theory and self-interaction-corrected 
density-functional theory using an exchange-only (Kohn-Sham) approximation to the density-functional
theory. Both theories correctly reproduce an antiferromagnetic arrangement of spins at very large 
separations. 

\begin{figure}
\epsfig{file=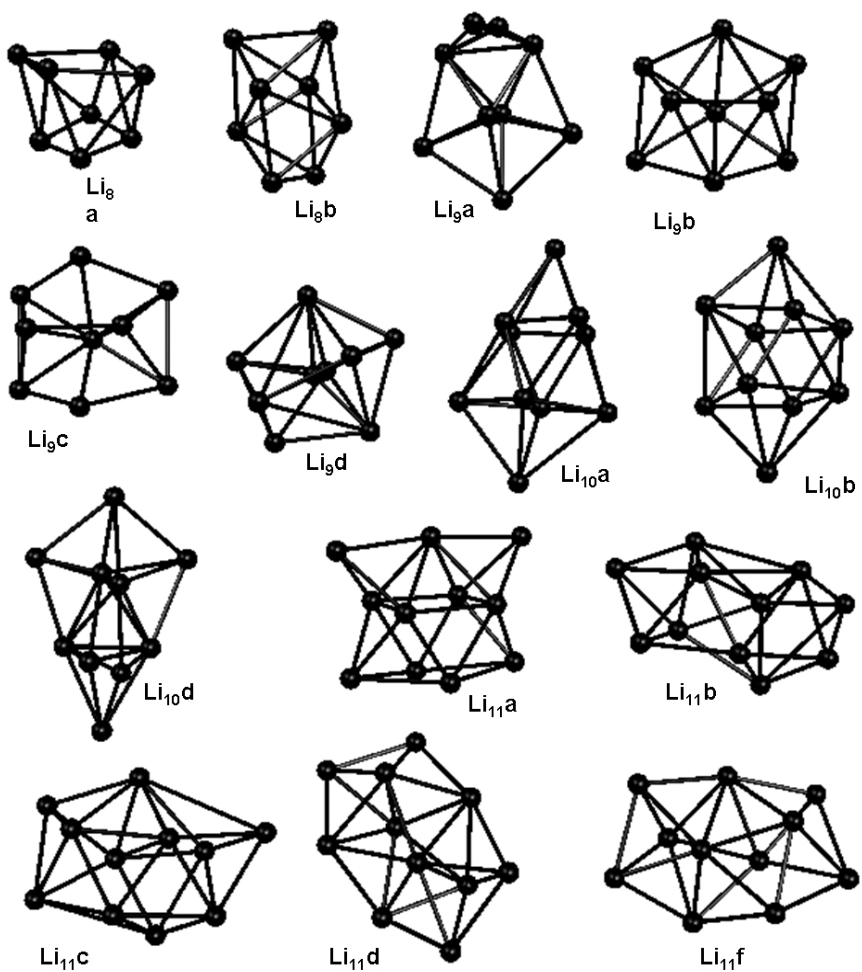,width=14cm,clip=true}
\caption{\label{fig:set1} 
The structures of lithium clusters in size range 8-11. Note that for a few sizes 
the labeling is not sequential.
(See text for more details).}
\end{figure}

\begin{figure}
\epsfig{file=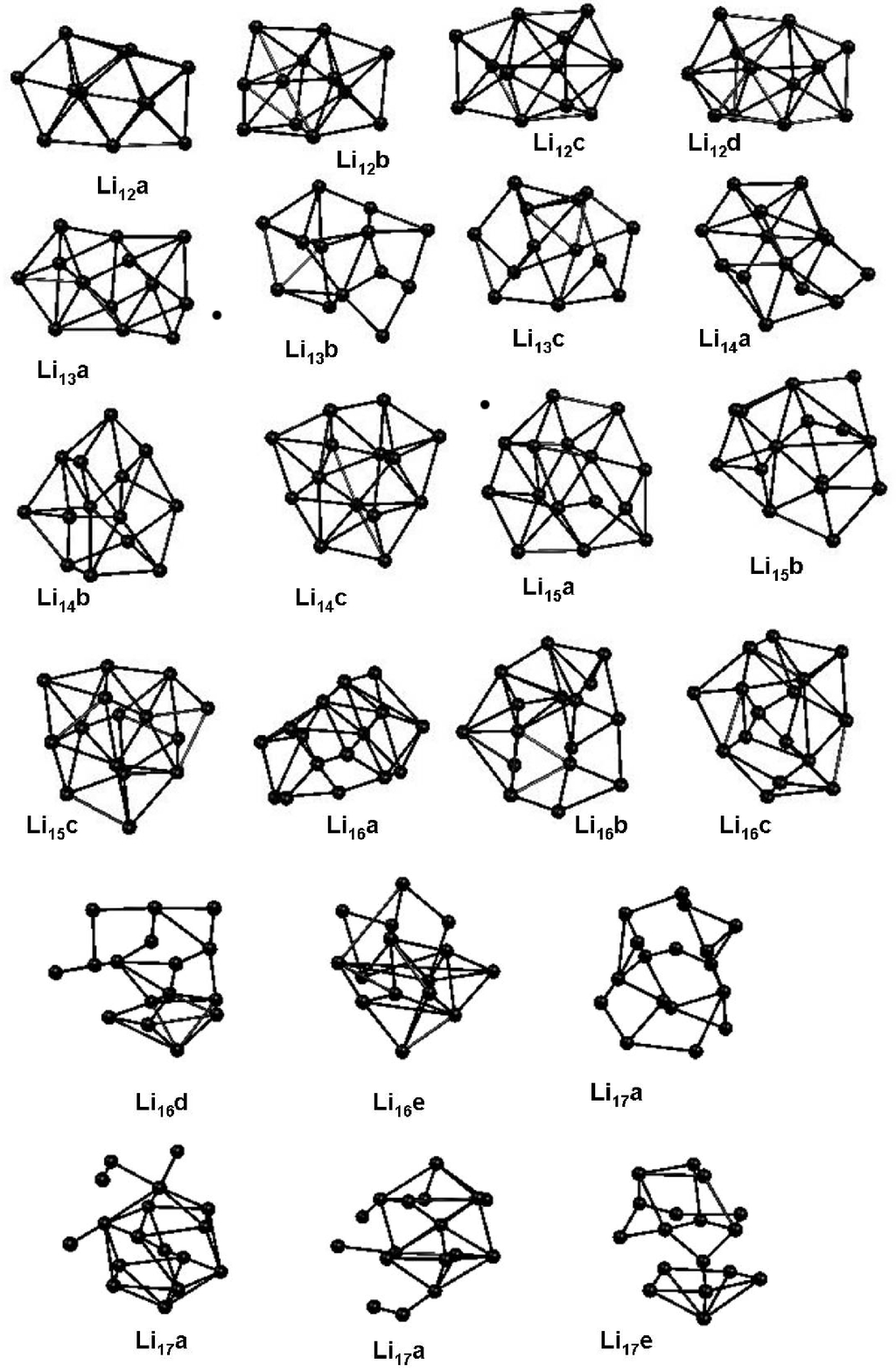,width=14cm,clip=true}
\caption{\label{fig:set2} 
The structures of lithium clusters in size range 12-16
(See text for more details).}
\end{figure}

\section{Results}
  The optimized structures of the Li clusters in the size range 8-17 are presented in Fig. 1 and 2.
Larger clusters are not presented as their structures are too complex to present and are not visually 
very informative.
Primary results of our calculations are presented in 
Table 3-5 and Figures 3 and 4.
With respect to binding energies we find that the
binding energy per atom increases slowly and monotonically as a function of cluster size. None of the clusters show
uncharacteristically large binding energies in this size regime.  The energy gaps, which are often indicators of cluster
reactivity and chemical stability, show some reasonably large variations. For example for N=20, all of the clusters have 
binding energies in the range of 1.16 eV/atom. However, one of the structures has a HOMO-LUMO gap that is 1.5 times larger
than the other two structures (Cf. Fig. ~\ref{fig:gap}). This is a case where the relative reactivities of the 
clusters could be the dominant 
factor in determining which cluster is actually being experimentally observed. However, even for this case the polarizability
tensors only differ at the level of a 1-3 percent which is small compared to the differences between
the experimental and theoretical results.
The dipole moments are also included in the tables. As expected for a nearly free-electron cluster the 
dipole moments are nearly negligible.  Our data suggests that issues related to dipole moments would be most likely
to be important for N=11, N=12, N=16 and N=19. The anisotropy of the dipole moment
may be determined from the {\em eigenvalues} of the polarizability tensor (labelled 
$\alpha_{xx}$, $\alpha_{yy}$, and $\alpha_{zz}$ in the tables. We emphasize here 
that these are the eigenvalues of the polarizability tensor and not only
the diagonal elements of the polarizability tensor.
\begin{table}[!h]
\begin{center}
\caption{Calculated results as a function of cluster size and conformation for N=12 to N=17. Results
include moment, $\mu$ in Bohr magnetons, binding energy (B.E.) in eV, average and principle 
polarizabilities ($\langle \alpha \rangle$, 
$\alpha_{xx}$,
$\alpha_{yy}$,
$\alpha_{zz}$ in \AA$^3$, the dipol moment ($p_x, p_y, p_z$) in atomic units and the HOMO ($\epsilon_H$) 
and LUMO ($\epsilon_L$)
energies in eV and HOMO-LUMO gap ($\Delta$) in eV respectively. 
The 
anisotropy of the polarizability tensor may be determined from the principle
polarizabilities (See Eq. 7 Ref.~\cite{mrp013}).
}
\begin{tabular}{|c|c|c|c|ccc|ccc|ccc|} \hline 
N & $\mu$ & B.E. & $\langle \alpha \rangle$&  $\alpha_{xx}$& $\alpha_{yy}$&$\alpha_{zz}$ & $p_x$& $p_y$& $p_z$&$\epsilon_H$&$\epsilon_L$&$\Delta$\\
\hline \hline
12-a&  0&     -1.01& 123.94& 164.68& 120.32&  86.82&  -0.01&   0.00&   0.00&  -2.89&  -2.33&   0.56\\
12-b&  0&     -1.03& 122.53& 162.02& 118.58&  86.98&  -0.08&  -0.06&   0.04&  -2.88&  -2.34&   0.54\\
12-c&  0&     -1.03& 122.58& 162.11& 118.54&  87.09&  -0.10&   0.01&  -0.03&  -2.88&  -2.34&   0.54\\
12-d&  0&     -1.03& 122.47& 161.97& 118.52&  86.91&   0.10&   0.00&  -0.05&  -2.88&  -2.34&   0.54\\
12-e&  0&     -1.03& 122.32& 161.73& 118.13&  87.11&   0.10&  -0.03&   0.07&  -2.87&  -2.32&   0.55\\
\hline
13-a&  1&     -1.03& 133.12& 171.89& 129.80&  97.67&   0.04&   0.01&  -0.01&  -2.67&  -2.29&   0.38\\
13-b&  1&     -1.04& 134.47& 167.48& 145.86&  90.08&   0.02&   0.01&  -0.02&  -2.78&  -2.29&   0.49\\
13-c&  1&     -1.04& 133.16& 165.51& 142.63&  91.34&   0.01&   0.00&   0.01&  -2.83&  -2.41&   0.43\\
\hline
14-a&  0&     -1.05& 142.63& 172.56& 165.80&  89.53&   0.00&   0.00&   0.00&  -3.02&  -2.18&   0.84\\
14-b&  0&     -1.07& 142.72& 169.07& 167.87&  91.22&   0.08&   0.01&   0.08&  -3.03&  -2.24&   0.78\\
14-c&  0&     -1.07& 142.72& 169.07& 167.87&  91.22&   0.08&   0.01&   0.08&  -3.02&  -2.24&   0.79\\
\hline
15-a&  1&     -1.05& 153.60& 184.89& 171.97& 103.94&  -0.06&  -0.03&   0.00&  -2.66&  -2.32&   0.35\\
15-b&  1&     -1.07& 151.09& 177.47& 168.33& 107.46&   0.04&   0.00&  -0.02&  -2.77&  -2.36&   0.41\\
15-c&  1&     -1.07& 151.04& 177.12& 168.36& 107.63&  -0.04&  -0.03&   0.00&  -2.78&  -2.36&   0.41\\
\hline
16-a&  0&     -1.08& 155.31& 179.89& 168.75& 117.29&  -0.06&  -0.02&  -0.02&  -2.76&  -2.39&   0.37\\
16-b&  0&     -1.08& 154.75& 178.71& 168.81& 116.75&  -0.25&   0.20&  -0.30&  -2.67&  -2.34&   0.33\\
16-c&  0&     -1.08& 156.68& 180.21& 171.11& 118.72&  -0.19&  -0.16&   0.04&  -2.65&  -2.42&   0.23\\
16-d&  0&     -1.08& 154.59& 179.97& 164.62& 119.20&  -0.31&  -0.06&   0.25&  -2.71&  -2.40&   0.32\\
16-e&  0&     -1.09& 154.46& 179.55& 165.82& 118.00&  -0.30&  -0.03&  -0.15&  -2.73&  -2.39&   0.34\\
\hline
17-a&  3&     -1.11& 156.15& 172.14& 169.58& 126.74&   0.05&   0.01&   0.02&  -2.78&  -2.53&   0.26\\
17-b&  3&     -1.11& 155.86& 171.82& 169.34& 126.41&  -0.05&  -0.04&   0.00&  -2.78&  -2.53&   0.26\\
17-c&  1&     -1.11& 155.98& 172.96& 170.32& 124.65&   0.06&   0.04&   0.02&  -2.81&  -2.45&   0.36\\
17-d&  3&     -1.13& 156.72& 171.86& 149.21& 149.09&   0.00&   0.00&   0.00&  -2.78&  -2.43&   0.35\\
\hline
\hline \hline
\end{tabular}
\end{center}
\end{table}

Simple models for shell closings in electronic systems have been proposed. These models are often based on
jellium descriptions but other potentials such as harmonic-oscillators and central-field models have also been
employed to predict shell closings. For the size regimes discussed here, the spherical jellium model would predict 
shell closings for N=8 and N=20. On the other hand, a central field model would predict shell closing for N=10, N=18 
and N=20. The asphericities that appear in the polarizability tensors for the case of N=10 and N=18 argue strongly
against an approximate central-field based description of Li clusters in this range. However, for N=20 and N=8 deviations
from sphericity are found to be only 2 and 12 percent respectively. The jellium like behavior is particularly striking
at N=20 and departures from sphericity at N=19 and N=21 are also quite pronounced. In comparison to N=19 and N=21 the 
HOMO-LUMO gap for the N=20 clusters is also large which again indicates a jellium-like behavior has emerged for clusters
in this size regime. For most of the clusters studied, the average polarizability did not show significant changes as
a function of conformation. However, for N=9 there is a 7 percent variation in average polarizability as a function of 
conformation. 
\begin{table*}
\begin{center}
\caption{Calculated results as a function of cluster size and conformation for N=18 to N=22. Results
include moment, $\mu$ in Bohr magnetons, binding energy (B.E.) in eV, average and principle 
polarizabilities ($\langle \alpha \rangle$, 
$\alpha_{xx}$,
$\alpha_{yy}$,
$\alpha_{zz}$ in \AA$^3$, the dipole moment ($p_x, p_y, p_z$) in atomic units and the HOMO ($\epsilon_H$) 
and LUMO ($\epsilon_L$) energies 
in eV and HOMO-LUMO gap ($\Delta$) in eV respectively. 
The 
anisotropy of the polarizability tensor may be determined from the principle
polarizabilities (See Eq. 7 Ref.~\cite{mrp013}).
}
\begin{tabular}{|c|c|c|c|ccc|ccc|ccc|} \hline 
N & $\mu$ & B.E. (eV)& $\langle \alpha \rangle$&  $\alpha_{xx}$& $\alpha_{yy}$&$\alpha_{zz}$ & $p_x$& $p_y$& $p_z$&$\epsilon_H$&$\epsilon_L$&$\Delta$\\
\hline \hline
18-a&  0&     -1.12& 165.36& 182.64& 176.66& 136.79&   0.08&   0.07&  -0.01&  -2.72&  -2.48&   0.24\\
18-d&  0&     -1.11& 165.82& 177.73& 175.19& 144.55&  -0.06&   0.02&   0.00&  -2.63&  -2.62&   0.01\\
18-e&  0&     -1.11& 162.84& 183.68& 168.58& 136.25&  -0.08&   0.00&   0.00&  -2.79&  -2.57&   0.21\\
18-f&  0&     -1.10& 157.79& 173.56& 162.22& 137.60&   0.18&   0.00&   0.00&  -3.04&  -2.70&   0.35\\
\hline
19-a&  1&     -1.14& 175.95& 186.30& 179.68& 161.86&   0.07&   0.07&   0.00&  -2.83&  -2.38&   0.45\\
19-c&  1&     -1.12& 175.25& 185.73& 185.20& 154.82&  -0.01&  -0.01&   0.00&  -3.02&  -2.52&   0.51\\
19-d&  1&     -1.12& 175.43& 185.57& 185.30& 155.42&   0.00&   0.00&   0.00&  -3.02&  -2.52&   0.50\\
19-e&  1&     -1.11& 171.09& 181.11& 172.15& 160.03&  -0.16&  -0.03&   0.00&  -3.06&  -2.72&   0.35\\
\hline
20-a&  0&     -1.16& 183.22& 184.56& 183.47& 181.63&   0.02&   0.00&   0.02&  -2.83&  -2.21&   0.62\\
20-b&  0&     -1.16& 181.62& 183.07& 181.96& 179.83&   0.03&   0.00&   0.02&  -2.83&  -2.21&   0.62\\
20-e&  0&     -1.16& 182.82& 184.71& 184.52& 179.23&   0.00&   0.00&   0.00&  -2.99&  -2.10&   0.89\\
20-c&  0&     -1.16& 181.17& 182.79& 181.62& 179.11&   0.04&   0.02&  -0.03&  -2.83&  -2.20&   0.63\\
20-d&  0&     -1.16& 180.87& 182.29& 181.10& 179.21&  -0.02&   0.00&   0.01&  -2.83&  -2.20&   0.63\\
\hline
21-a&  1&     -1.16& 189.70& 210.47& 179.49& 179.13&   0.00&   0.00&   0.00&  -2.68&  -2.39&   0.28\\
21-d&  1&     -1.16& 192.43& 214.14& 182.04& 181.13&  -0.07&   0.00&   0.01&  -2.59&  -2.25&   0.34\\
21-f&  1&     -1.16& 193.01& 211.57& 185.58& 181.89&   0.03&   0.00&   0.00&  -2.62&  -2.31&   0.31\\
\hline
22-a&  0&     -1.16& 204.14& 245.17& 183.83& 183.44&   0.00&   0.00&   0.00&  -2.78&  -2.41&   0.37\\
22-c&  0&     -1.16& 202.98& 243.81& 184.15& 180.99&   0.03&   0.00&   0.00&  -2.77&  -2.38&   0.40\\
22-f&  0&     -1.15& 207.80& 252.69& 185.72& 184.98&   0.01&   0.00&   0.00&  -2.76&  -2.42&   0.35\\
22-g&  0&     -1.15& 205.79& 252.25& 183.53& 181.61&  -0.07&  -0.01&  -0.01&  -2.77&  -2.40&   0.37\\
22-h&  0&     -1.15& 207.67& 253.03& 185.31& 184.66&   0.00&   0.00&   0.00&  -2.76&  -2.42&   0.34\\
\hline \hline
\end{tabular}
\end{center}
\end{table*}

\section{Summary and Conclusions}
In this paper we have discussed a relatively automated procedure for investigating polarizabilities
of clusters  as a 
function of size and conformation. This procedure has been tested  on Lithium clusters in the N=8-22 range
due to availability of experimental data. While the binding energy per atom and average polarizability
show very little nonmonotonic behavior in this range, there are significant departures from isotropic
polarizabilities in these clusters. As expected from jellium models it is only for N=20, the 
polarizability tensor is found to be significantly isotropic.   
The calculated polarizabilities show smoother decrease with increase in cluster size than the
experimental values.
The spin polarized calculations 
indicate that all even size clusters as expected have zero moments (no unpaired electron). However, a
few odd size clusters are found to show higher ( 3 $\mu_{\beta}$) spin states.

\begin{figure}
\epsfig{file=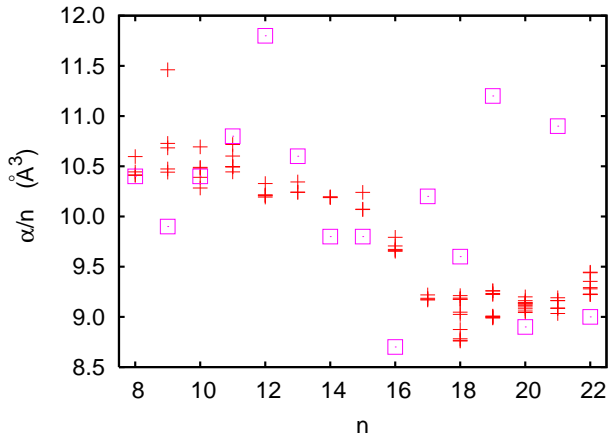,width=8.5cm,clip=true}
\caption{\label{fig:pol} (Color online) 
Static dipole polarizability per atom (\AAA) of lithium clusters as a 
function of number of constituent atoms. The polarizability is 
presented for a number of isomers for each size. The squares represent 
experimental values.} 
\end{figure}

\begin{figure}
\epsfig{file=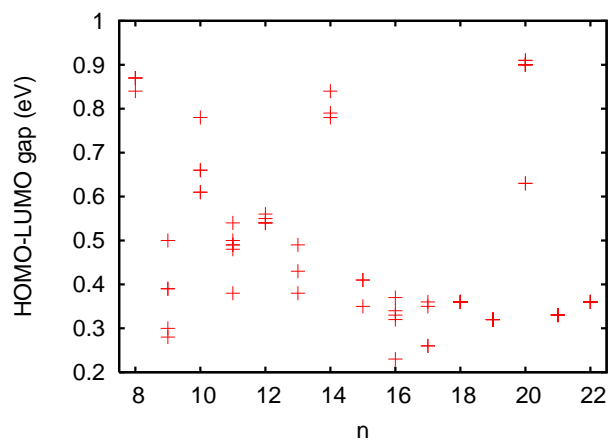,width=8.5cm,clip=true}
\caption{\label{fig:gap} 
The HOMO-LUMO gap (eV) of lithium clusters as a 
function of number of constituent atoms. }
\end{figure}

\section{Acknowledgements} 
We thank Dr. P. Calaminici and Dr. G. Maroulis for encouraging us to complete 
the present work for this issue.  This work is supported in part by
the National Science Foundation through CREST grant,
by the University of Texas at El Paso (UTEP startup funds) and  in part
by the Office of Naval Research, directly (ONR 05PR07548-00) and through 
the Naval Research Laboratory.   Computational support was provided by
the DoD High Performance Computing Modernization Office.

\end{document}